\documentclass[conference]{IEEEtran}
\IEEEoverridecommandlockouts
\usepackage{cite}
\usepackage{amsmath,amssymb,amsfonts}
\usepackage{algorithmic}
\usepackage{graphicx}
\usepackage{textcomp}
\usepackage{xcolor}
\usepackage{booktabs}
\usepackage{arydshln} 
\usepackage{multirow}
\usepackage{multicol}
\usepackage{colortbl}
\usepackage{hyperref}
\usepackage{tikz}

\newcommand{\etal}{\textit{et al.}}
\newcommand\constructosum[3]{%
    \begin{tikzpicture}[baseline=(char.base), inner sep=0, outer sep=0]
        \draw (#1,0) circle (#2); 
        \node (char) at (0,0) {$#3\sum$}; 
    \end{tikzpicture}%
}

\newcommand{\osum}{\mathop{\mathchoice
        {\constructosum{-0.3ex}{0.1}{\displaystyle}}
        {\constructosum{-0.3ex}{0.06}{\textstyle}}
        {\constructosum{-0.2ex}{0.04}{\scriptstyle}}
        {\constructosum{-0.15ex}{0.03}{\scriptscriptstyle}}
    }\displaylimits
}
\def\BibTeX{{\rm B\kern-.05em{\sc i\kern-.025em b}\kern-.08em
    T\kern-.1667em\lower.7ex\hbox{E}\kern-.125emX}}
\begin{document}

\title{Flemme: A Flexible and Modular Learning Platform for Medical Images}

\author{\IEEEauthorblockN{Guoqing Zhang\textsuperscript{1, 2}, Jingyun Yang\textsuperscript{1}, Yang Li\textsuperscript{1}}
\IEEEauthorblockA{\textsuperscript{1}\textit{Tsinghua Shenzhen International Graduate School, Tsinghua University} \\
\textsuperscript{2}\textit{Peng Cheng Laboratory}\\
Shenzhen, China \\
zhanggq21@mails.tsinghua.edu.cn}
\vspace{-2em}}
\maketitle
%
\begin{abstract}
As the rapid development of computer vision and the emergence of powerful network backbones and architectures, the application of deep learning in medical imaging has become increasingly significant. Unlike natural images, medical images lack huge volumes of data but feature more modalities, making it difficult to train a general model that has satisfactory performance across various datasets. In practice, practitioners often suffer from manually creating and testing models combining independent backbones and architectures, which is a laborious and time-consuming process. We propose \textit{Flemme}, a \textit{FLE}xible and \textit{M}odular learning platform for \textit{ME}dical images. Our platform separates encoders from the model architectures so that different models can be constructed via various combinations of supported encoders and architectures. We construct encoders using building blocks based on convolution, transformer, and state-space model (SSM) to process both 2D and 3D image patches. A base architecture is implemented following an encoder-decoder style, with several derived architectures for image segmentation, reconstruction, and generation tasks. In addition, we propose a general hierarchical architecture incorporating a pyramid loss to optimize and fuse vertical features. Experiments demonstrate that this simple design leads to an average improvement of 5.60\% in Dice score and 7.81\% in mean interaction of units (mIoU) for segmentation models, as well as an enhancement of 5.57\% in peak signal-to-noise ratio (PSNR) and 8.22\% in structural similarity (SSIM) for reconstruction models. We further utilize \textit{Flemme} as an analytical tool to assess the effectiveness and efficiency of various encoders across different tasks. Code is available at \href{https://github.com/wlsdzyzl/flemme}{https://github.com/wlsdzyzl/flemme}.
\end{abstract}


\begin{figure*}[h]
	\centering
	\includegraphics[width=2\columnwidth, angle=0]{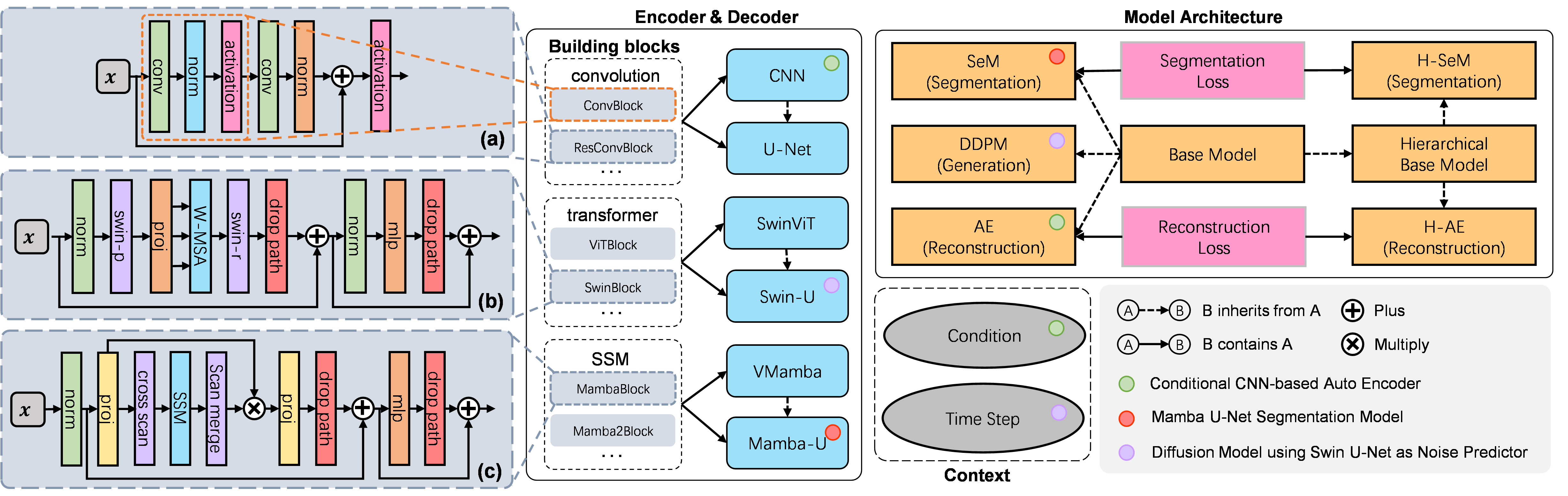}
        \vspace{-0.5em}
	\caption{A semantic overview of Flemme. The left box gives 3 examples of building blocks based on convolution, transformer, and SSM. Encoders and architectures are shown in the middle and right blocks. Different models can be constructed efficiently from various combinations of encoders and architectures labeled with the corresponding colors.}
	\label{fig:overview}
 \vspace{-1em}
\end{figure*}
\section{Introduction}

Since AlexNet \cite{krizhevsky2012imagenet} competed in the ImageNet Large Scale Visual Recognition Challenge \cite{deng2009imagenet}, convolutional neural networks (CNNs) have become dominant in computer vision. In particular, the most preferred choice in medical imaging is U-Net \cite{ronneberger2015u} which uses skip connections to enhance feature fusion across different time stages. However, the limitation of convolution lies in its focus on local feature extraction, lacking the ability to explore long-range dependencies. In recent years, following the immense success of transformers in natural language processing (NLP) \cite{vaswani2017attention,devlin2018bert}, treating images as sequences has garnered significant attention. Dosovitskiy \etal \cite{dosovitskiy2020image} propose vision transformer (ViT) to encode sequences of image patches. While ViT outperforms CNN-based models \cite{he2016deep} in image recognition, the quadratic computational complexity of the multi-head self-attention (MSA) mechanism makes it difficult to scale to larger images. Swin Transformer \cite{liu2021swin} introduces a shifted window attention mechanism to reduce computational complexity and significantly improves the applicability of transformers for vision tasks. On the other hand, state space models (SSMs) have also made significant advances in sequence modeling. Gu \etal \cite{gu2023mamba} propose a general sequence model backbone named Mamba with a linear time complexity, allowing for consideration of a much longer range of dependencies compared to transformers. Zhu \etal \cite{zhu2024vision} and Liu \etal \cite{liu2024vmambavisualstatespace} process image patches with vision mamba blocks and introduced bidirectional and cross-scan strategies, respectively, to effectively integrate vision information from different directions. 

Although new methods continue to emerge, several aspects remain worthy of exploration. Firstly, model performance depends highly on actual training techniques and deployment. For instance, Liu \etal \cite{liu2022convnet} demonstrates that CNNs can be "modernized" to achieve performance compatible with vision transformers. In real-world applications, the improvement in model performance needs to be balanced against the computational costs of training and inference. Secondly, beyond classification and segmentation, powerful model architectures have been proposed for image reconstruction and generation, such as variational auto-encoder (VAE) \cite{kingma2013auto} and diffusion models \cite{ho2020denoising, song2020denoising}. Applying the advanced backbones to these architectures for medical images is promising and worth anticipating. Thirdly, unlike natural images, medical images often lack large-scale, high-quality annotations but encompass a wide range of modalities, making it hard to train a general model that performs well across various medical image datasets. In practice, researchers and engineers often need to manually build and test models with multiple backbones and architectures for specific tasks. However, combining independent methods can be labor-intensive and hard to analyze.

Given the aforementioned challenges, the lack of a universal platform that supports the fast construction of diverse models significantly adds barriers to adopting the latest technologies and increases research duration. We propose \textbf{Flemme}, a \textbf{FLE}xible, and \textbf{M}odular learning platform for \textbf{ME}dical images. Our platform decouples the encoder from the model architecture, enabling the fast construction of models by combining different encoders and architectures. We employ convolution, vision transformer, and SSM as backbones to create encoders and decoders for both 2D and 3D images. Our base model architecture follows an encoder-decoder structure, with further derivations for segmentation, reconstruction, and generation tasks. In addition to fusing horizontal features of the same scale among different stages like U-Net \cite{ronneberger2015u}, we propose a general hierarchical framework for vertical feature fusion across different scales. To summarize, Flemme is distinguished by the following features:
\begin{itemize}
  \item We adopt a modular design with state-of-the-art encoders and architectures for fast model construction.
  \vspace{0.25em}
  \item We implement a novel backbone-agnostic hierarchical architecture combining a pyramid loss for generic vertical feature fusion and optimization.
  \vspace{0.25em}
  \item We support flexible context encoding and allow extensions of new modules for more data types and tasks.
\end{itemize}

We demonstrate the applications of the proposed platform on various medical image datasets. For all supported models, we can set hyper-parameters such as network depth, normalization, and training strategies in the same manner. Benefiting from this advantage, we conduct extensive experiments with various encoders on different tasks to fairly compare and analyze the performance, as well as the time and memory complexity. We hope our work will set new benchmarks and serve as a valuable research tool for future investigations into the potential of convolutions, transformers, and SSMs in medical imaging.


\section{Methods}
An overview of the proposed platform is presented in Fig.~\ref{fig:overview}, in which we show examples of how to construct different models with various encoders and architectures for different tasks. Our platform mainly consists of three modules: encoder \& decoder, context embedding, and model architectures, of which context embedding is optional. The remainder of this section is organized as follows: we give mathematical formulations of various building blocks and derived encoders in Sec.~\ref{sec:encoder}. Sec.~\ref{sec:con_encoding} introduces our context encoding strategies. Sec.\ref{sec:model_archi} elaborates supported architectures for different tasks. 
\begin{figure}[h]
	\centering
	\includegraphics[width=\columnwidth, angle=0]{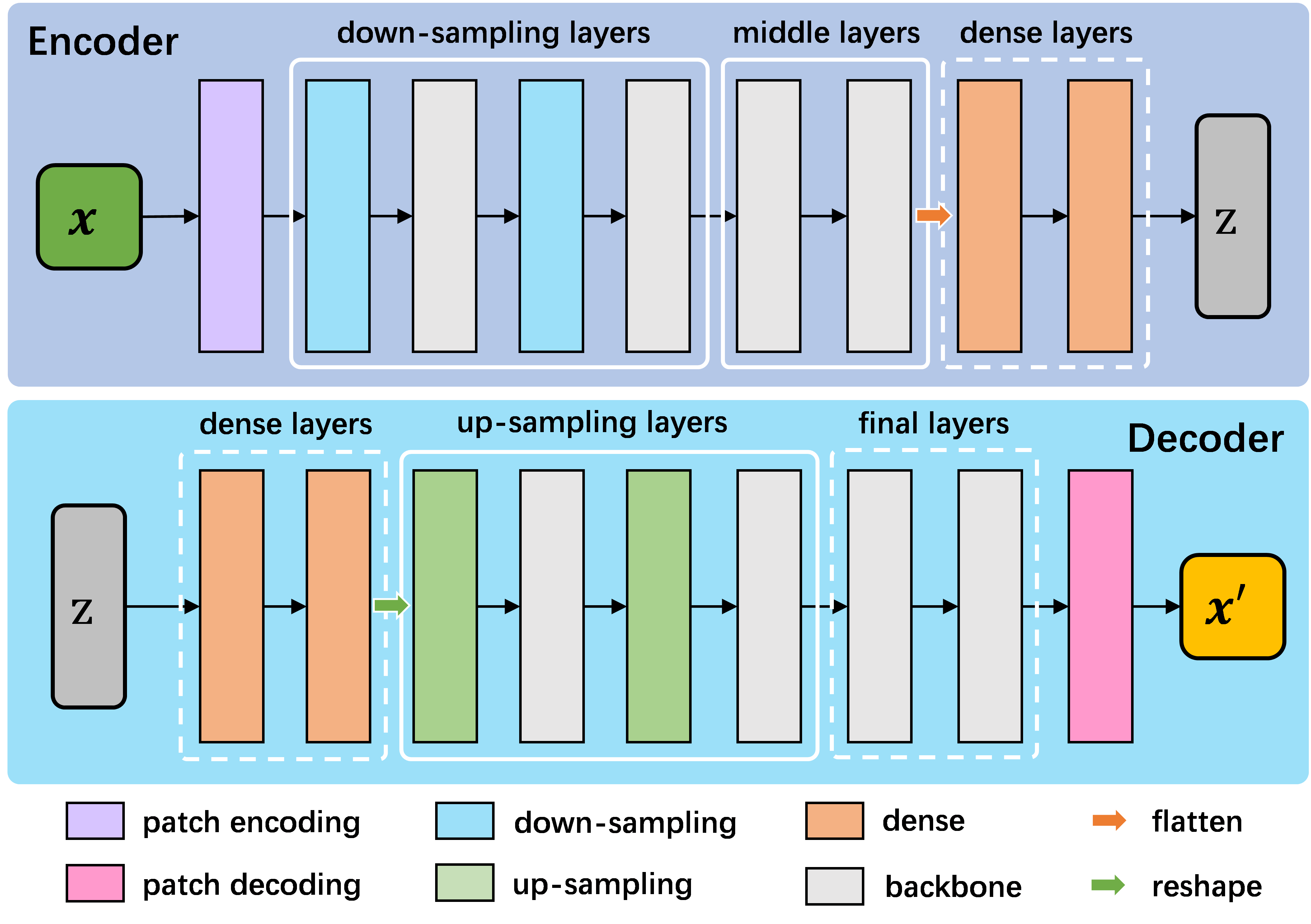}
        \vspace{-1.5em}
	\caption{Pipelines of encoder and decoder. Components enclosed in dotted boxes indicate optional elements.}
    \label{fig:encoder-decoder}
\end{figure}
\begin{figure*}[ht]
	\centering
	\includegraphics[width=2\columnwidth, angle=0]{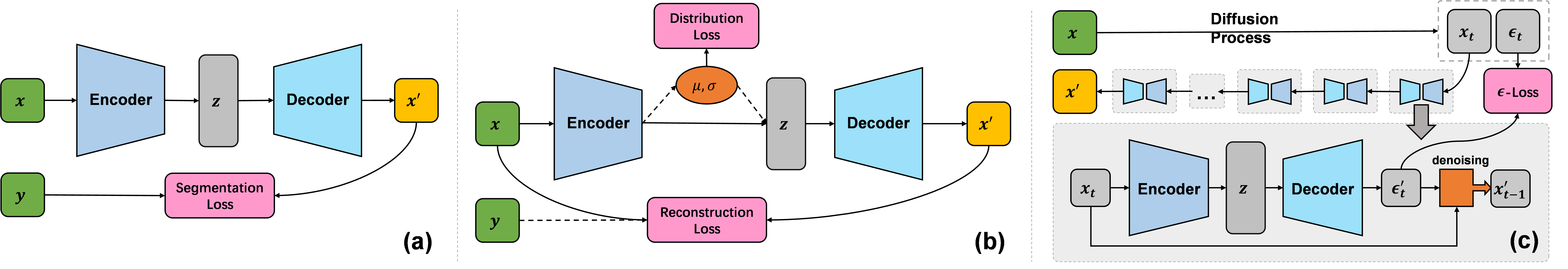}
        \vspace{-0.5em}
	\caption{Illustration of supported architectures: (a) SeM, (b) AE, (d) DDPM. The dashed lines indicate optional paths.}
        \label{fig:supported_archis}
 \vspace{-1em}
\end{figure*}
\subsection{Encoder and Decoder}
\label{sec:encoder}
We implement all encoders and the corresponding decoders following pipelines illustrated in Fig.~\ref{fig:encoder-decoder}. An encoder consists of a patch encoding block, down-sampling layers, middle layers, and optional densely connected layers. Patch encoding block transfers input images into patch embeddings. A basic element of down-sampling layers contains a down-sampling block and a building block to compress image patches into smaller but deeper feature maps. Middle layers contain only building blocks for further information extraction. Finally, we can choose to flatten the feature maps and obtain 1-d vector embeddings through dense layers, or directly pass the feature maps into the decoder. 

The decoder performs the reverse process of the encoder. It takes the latent features as input and reshapes them into feature maps if the inputs are 1D vector embeddings. The up-sampling layers consist of up-sampling blocks and building blocks for feature expansion and localization. Final layers are optional for further feature enhancement. A patch decoding block is then used to transfer feature maps into pixel space.

For convolution-based encoders, we use convolution to implement patch encoding and down-sampling, and transposed convolution for patch decoding and up-sampling. For encoders and decoders using transformers or SSMs as backbones, patch encoding, and decoding involve additional feature permutation to facilitate sequence modeling, with down-sampling and up-sampling implemented through patch merging and patch expansion operations as introduced in \cite{liu2021swin}. The main difference between various encoders lies in the employed building blocks, which are detailed in the remainder of this section.

\subsubsection{Convolution} We introduce two types of building blocks based on convolution: ConvBlock and ResConvBlock. A \textbf{ConvBlock} consists of a convolution, a normalization layer, and an activation function. Formally, we consider a forward pass of ConvBlock defined as:
\begin{equation}
\begin{split}
    \mathcal{F}_{C}(x) =  
                        \operatorname{RL}(\operatorname{GN}(\operatorname{Conv}(x)))
\end{split}
\end{equation}
where $\operatorname{RL}$ and $\operatorname{GN}$ denote the rectified linear unit (ReLU) and group normalization \cite{wu2018group}, which are the default choices of activation and normalization functions. Two MLPs, named Gate and Bias, are optional to introduce a gate-bias mechanism if there is a context vector embedding input $t$ \cite{grathwohl2018ffjord}: 
\begin{equation}
\begin{split}
    \mathcal{F}_{C}(x, t) =  \mathcal{F}_{C}(x) \cdot \operatorname{Gate}(t) + \operatorname{Bias}(t).
\end{split}
\end{equation}
As illustrated in Fig.~\ref{fig:overview} (a), \textbf{ResConvBlock} contains two ConvBlocks and introduces a skip connection for residual learning: 
\begin{equation}
\begin{split}
\label{eqn:res}
    \mathcal{F}_{RC}(x, t) = 
                          \operatorname{RL}\left(x + \mathcal{F}_{C}^2( \mathcal{F}_{C}^1 (x) + \operatorname{Proj}(t))\right).
\end{split}
\end{equation}
where $\operatorname{Proj}$ is a MLP for context embedding projection, $\mathcal{F}_{C}^1$ and $\mathcal{F}_{C}^2$ are the first and second ConvBlock, respectively. Note that $\mathcal{F}_{C}^2$ in ResConvBlock follows a similar manner to \cite{he2016deep} and has no activation function.
\subsubsection{Transformer} Following Dosovitskiy \etal \cite{dosovitskiy2020image}, we implement ViTBlock
and further introduce the shifted-window strategy \cite{liu2021swin} to construct \textbf{SwinBlock} for time and space complexity reduction. Specifically, we partition image patches into smaller windows, and a window-based MSA (W-MSA) is employed to capture dependencies only among patches within the same window. Relative position encoding for both 2D and 3D windows are used to further improve the modeling capability. By shifting the windows at different stages, overlaps among windows are created to increase the receptive field.  Fig~\ref{fig:overview} (b) gives an intuitive overview of SwinBlock that can be formulated as the following:
\begin{equation}
\begin{split}
    &z = \operatorname{Drop}(\operatorname{W-MSA}( \operatorname{LN}(x)) + x\\,
    &\mathcal{F}_{S}(x) =  \operatorname{Drop}(\operatorname{MLP}( \operatorname{LN}(z))) + z,
\end{split}
\end{equation}
where $\operatorname{LN}$ and $\operatorname{Drop}$ refers to layer normalization \cite{ba2016layer} and drop path \cite{larsson2016fractalnet}, $\operatorname{W-MSA}$ represents a series of operations, including patch shift, window partition, applying window-based MSA and the reversed operations. Similar to ConvBlock, we also introduce \textbf{ResSwinBlock} $\mathcal{F}_{RS}$, by simply replacing $\mathcal{F}_{C}(x)$ with $\mathcal{F}_{S}(x)$ in Eqn.~(\ref{eqn:res}).
\subsubsection{State-Space Model} SSM is a linear time-invariant system to map a 1-d sequence $x(t)$ to $y(t)$ through a hidden state $h(t)$. We adopt Mamba \cite{gu2023mamba, dao2024transformers} as the implementation of SSM to construct \textbf{MambaBlock}. In addition, we traverse the 2D and 3D image patches in a cross-scan manner and jointly model the obtained sequences to enhance the spatial awareness of Mamba. As depicted in Fig.~\ref{fig:overview} (c), MambaBlock processes image patches as follows:
\begin{equation}
\begin{split}
    & w =  \operatorname{Proj}(\operatorname{LN}(x)),\\
    &z = \operatorname{Drop}(\operatorname{Proj}^{-1}(\operatorname{SSM}(w) \cdot \operatorname{MLP}(w))) + x,\\
    &\mathcal{F}_{M}(x) =  \operatorname{Drop}(\operatorname{MLP}( z)) + z,
\end{split}
\end{equation}
where $ \operatorname{Proj}$ projects image patch into inner space, $\operatorname{SSM}$ indicates multiple operations including cross scanning, sequence modeling with Mamba, and scan merging. \textbf{ResMambaBlock} $\mathcal{F}_{RM}$ are introduced by using $\mathcal{F}_{M}(x)$ as a replacement for $\mathcal{F}_{C}(x)$ in Eqn.~(\ref{eqn:res}).
\subsubsection{U-shaped Networks} 
\begin{figure}[h]
	\centering
	\includegraphics[width=\columnwidth, angle=0]{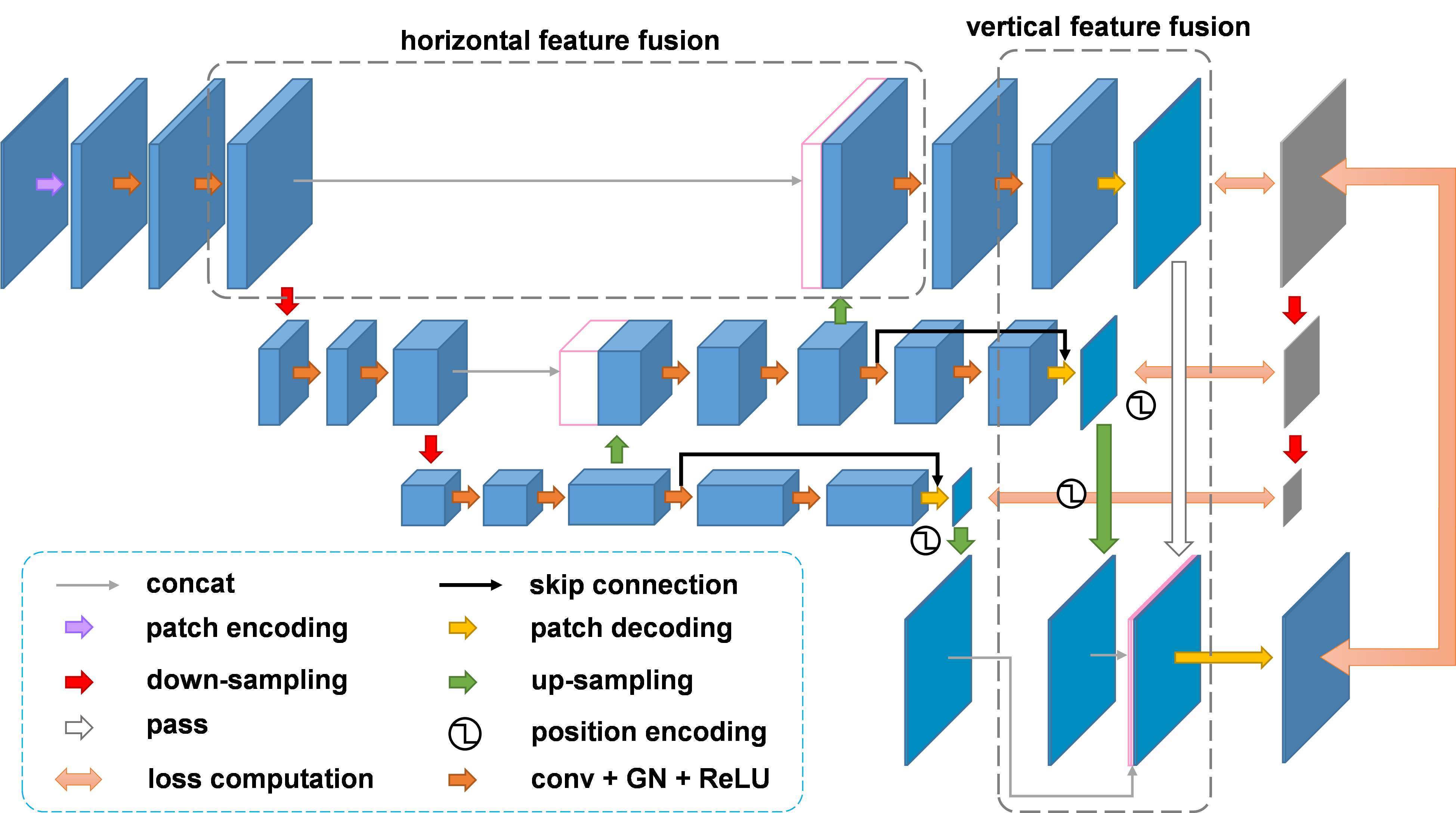}
        \vspace{-1.5em}
	\caption{A segmentation model constructed with Hierarchical SeM (H-SeM) and a U-shaped encoder using ConvBlock.}
        \label{fig:pyramid}
 \vspace{-0.5em}
\end{figure}

For all the aforementioned encoders, we also implement their U-shaped variants. The U-shaped encoder outputs feature maps from down-sampling layers, which are concatenated with the feature maps from up-sampling layers of corresponding spatial resolutions in the decoder. U-shaped networks have demonstrated superiority and become the de-facto standard in various imaging tasks \cite{ronneberger2015u, ho2020denoising, cao2022swin, chen2021transunet, wang2024mamba}. Fig.~\ref{fig:pyramid} gives an illustration of a U-shaped structure and horizontal feature integration.

Because the building blocks are capable of handling context embeddings, the encoder and decoder can also incorporate an additional context embedding as input. To summarize, the basic encoding and decoding processes can be formulated as follows:
\begin{equation}
\begin{split}
& \mathcal{E}(x, t) = \operatorname{Middle}(\operatorname{Down}(\mathcal{P}(x), t), t), \\
& \mathcal{D}(z, t) = \mathcal{P}^{-1}\left(\operatorname{Up}(z, t)\right),
\end{split}
\end{equation}
where $\mathcal{E}$ and $\mathcal{D}$ represent the encoder and decoder, $\mathcal{P}(\cdot)$ and $\mathcal{P}^{-1}(\cdot)$ refer to the patching encoding and decoding operations, $\operatorname{Down}$ and $\operatorname{Up}$ refer to the down-sampling and up-sampling layers, and $\operatorname{Middle}$ refers to the middle layers, respectively.
\subsection{Context embedding}
\label{sec:con_encoding}
Encoding contexts is necessary for conditional imaging tasks \cite{van2016conditional} and diffusion models  \cite{ho2020denoising,ho2022classifier}. Context refers to any additional input other than the input image, which can be a class label, a prompt image, or a time step index from diffusion process, We compute one-hot and sinusoidal position embeddings \cite{vaswani2017attention} for class label and time step index. When the context is an image, any of the encoders introduced in Sec.~\ref{sec:encoder} can be employed. 

\begin{table*}[h]
\caption{Evaluation of segmentation models. The best results are denoted in \textbf{Bold}. The preferred  result of each encoder with different architectures is indicated in \textit{Italics}. The second best performing encoder's preferred result is \underline{Underlined}. }
\vspace{-1.6em}
\label{tab:seg}
\renewcommand{\arraystretch}{1.1}

\begin{center}
\resizebox{2\columnwidth}{!}{
\begin{tabular}{l|l|cc|cc|cc|cc|cc|cc}
\bottomrule
\multirow{2}{*}{{Encoder}} & \multirow{2}{*}{{Archi}}  &     \multicolumn{2}{c|}{{CVC-ClinicDB}} &         \multicolumn{2}{c|}{{Echonet}} &  \multicolumn{2}{c|}{{ISIC}} &    \multicolumn{2}{c|}{{TN3K}} &        \multicolumn{2}{c|}{{BraTS21}}               &              \multicolumn{2}{c}{{ImageCAS}} \\ \cline{3-14}
    &   & {Dice}  & {mIoU}            & {Dice}                        & {mIoU}    &{Dice}   & {mIoU}   & {Dice}  & {mIoU}    & {Dice}   & {mIoU}    & {Dice}   & {mIoU}     \\ \hline
\multirow{2}{*}{{ResNet}}   &  SeM                 & \textit{0.8112} & \textit{0.7287}               & \textit{0.9193} & \textit{0.8538}                       & 0.8827                    & 0.8055                    & \textit{0.7105}                    & \textit{0.6048}                       & \textit{0.3091}                    & \textit{0.2813}                       & \textit{0.7076}                    & \textit{0.5497}                        \\
 & H-SeM                          & 0.8098         & 0.7278                      & 0.9183 & 0.8521                       & \textit{0.8836}                    & \textit{0.8087}                    & 0.6257                    & 0.5227                       & 0.2557                    & 0.2525                       & 0.7015                    & 0.5425                        \\
\hline
\multirow{2}{*}{{U-Net}} & SeM                         & 0.8457          & 0.7699                        & 0.9232 & 0.8604                       & 0.8864                    & 0.8113                    & 0.7344                    & 0.6329                       & 0.5008                    & 0.4262                       & 0.7450                     & 0.5961                        \\
 & H-SeM                            & \underline{\textit{0.8492}}           & \underline{\textit{0.7727}}                        & \textbf{0.9245} & \textbf{0.8626}                       & \textit{0.8915}                    & \textit{0.8184}                    & \underline{\textit{0.7395}}                    & \textbf{0.6449}                         & \textit{0.7984}                    & \textit{0.7182}                       & \underline{\textit{0.7595}}                    & \underline{\textit{0.6148}}                        \\
\hline
\multirow{2}{*}{{CAtten-U}} & SeM                           & 0.5431          & 0.4230                        &  0.9146 & 0.8464                       & 0.4121                   & 0.2836                    & 0.6096                  & 0.4864                      & -                    & -                       & -                     & -                        \\
 &H-SeM                            & \textit{0.8186}          & \textit{0.7401}                       & \textit{0.9231} & \textit{0.8600}                       & \textit{0.7834}                    & \textit{0.6858}                    & \textit{0.7208}                    & \textit{0.6179}                         & -                    & -                       & -                   & -                       \\
\hline
\multirow{2}{*}{{Swin-U}} & SeM                            & 0.6874          & 0.5800                        & 0.9125 & 0.8428                       & 0.8515                    & 0.7675                    & 0.6051                    & 0.4761                       & 0.8133                    & 0.7361                       & 0.7345                    & 0.5833                        \\
&H-SeM                         & \textit{0.8388}          & \textit{0.7572}                        & \textit{0.9182} & \textit{0.8521}                       & \underline{\textit{0.8980}}                     & \underline{\textit{0.8299}}                    & \textit{0.6729}                    & \textit{0.5525}                       & \underline{\textit{0.8406}}                    & \underline{\textit{0.7670}}                        & \textit{0.7475}                    & \textit{0.5997}                        \\
\hline
\multirow{2}{*}{{Mamba-U}}  & SeM                              & 0.8618          & 0.7937                        & 0.9225 & 0.8595                       & 0.8999                    & 0.8326                    & 0.7196                     & 0.6082                       & 0.8430                     & \textbf{0.7754}                       & 0.7644                    & 0.6215                        \\
& H-SeM                           & \textbf{0.8700}           & \textbf{0.8033}                        & \underline{\textit{0.9234}} & \underline{\textit{0.8609}}                       & \textbf{0.9050}                     & \textbf{0.8394}                    & \textbf{0.7506}                    & \underline{\textit{0.6472}}                       & \textbf{0.8450}                     & {0.7748}                       & \textbf{0.7678}                    & \textbf{0.6259}                       \\
\hline
\end{tabular}
}
\end{center}
\vspace{-2em}
\end{table*}

\subsection{Model Architecture}
\label{sec:model_archi}
\subsubsection{Encoder-decoder Architecture}
Our base architecture follows an encoder-decoder style with optional components for encoding context as mentioned in Sec.~\ref{sec:con_encoding}. A forward pass of based architecture is defined as:
\begin{equation}
\label{equ:forward}
\begin{split}
&z = \mathcal{E}\left(x + \mathcal{C}_e (c), \mathcal{T}(t)\right),\\
&\mathcal{M}(x, c, t) = \mathcal{D}\left( z + \mathcal{C}_d (c), \mathcal{T}(t )\right),
\end{split}
\end{equation}
where $\mathcal{C}_e$ and $\mathcal{C}_d$ are functions of encoding context $c$ for encoder $\mathcal{E}$ and decoder $\mathcal{D}$, $\mathcal{T}(\cdot)$ computes the sinusoidal positional embedding, $z$ is the latent embedding computed by the encoder, $c$ and $t$ are input contexts. Usually, $c$ is the input condition such as a prompt image or class label and $t$ is a time step index. Base architecture is not directly trainable due to the lack of loss functions and serves as a foundational structure for other architectures as illustrated in Fig.~\ref{fig:supported_archis}, including the Segmentation Model (SeM), the Auto-Encoder (AE) and the De-noising Diffusion Probabilistic Model (DDPM), for medical image segmentation, reconstruction, and generation, respectively. We explain these architectures in the remaining part of this section. To simplify the notation, we omit the optional input contexts in the following formulas to simplify the notation.
\subsubsection{Segmentation}As shown in Fig.~\ref{fig:supported_archis} (a), SeM is a simple extension of base architecture by introducing segmentation loss. It takes raw images as inputs and gives the predictions through a forward pass defined in Eqn.~(\ref{equ:forward}). Loss is computed over prediction and ground truth: 
\begin{equation}
\begin{split}
    \mathcal{L}_\text{SeM} = \mathcal{L}_\text{CE}(y', y) + \mathcal{L}_\text{Dice}(y', y),
\end{split}
\end{equation}
where $y'= \mathcal{M}(x)$ is the prediction, $y$ specifies the ground-truth. By default, segmentation loss $\mathcal{L}_\text{seg}$ is a combination of cross-entropy and Dice loss, denoted by $ \mathcal{L}_\text{CE}$ and $ \mathcal{L}_\text{Dice}$, respectively.

\subsubsection{Reconstruction}We implement auto-encoder (AE) for medical image restoration, which builds reconstruction loss over predictions and inputs, indicating an unsupervised representation learning:  
\begin{equation}
    \mathcal{L}_\text{AE} = \mathcal{L}_\text{MSE}(x', x),
\end{equation}
where $\mathcal{L}_\text{MSE}$ is the mean squared error, serving as the default reconstruction loss. As shown in Fig.~\ref{fig:supported_archis} (b), AE can be regularized by distribution loss to learn a more continuous latent representation and have a certain capability of generation \cite{kingma2013auto}.
The learning process of AE can also be supervised by a ground-truth target.

\subsubsection{Generation}Fig.~\ref{fig:supported_archis} (c) illustrates the pipeline of DDPM \cite{ho2020denoising}, in which base architecture is used to construct a noise predictor, denoted as $\epsilon$-model. In the diffusion process, we add random Gaussian noise to the input image to get $t$-th step noisy images, which can be formulated as the following:
\begin{equation}
    x_t = \alpha_t x +  \beta_t \epsilon,
\end{equation}
where $\alpha_t$ and $\beta_t$ are hyper-parameters related to noise schedules. The $\epsilon$-model takes noisy images as inputs and computes loss over the noise and prediction: 
\begin{equation}
    \mathcal{L}_\text{DDPM} =  \mathcal{L}_\text{MSE}(\epsilon', \epsilon),
\end{equation}
where $\epsilon' = \mathcal{M}_\epsilon(x_t, t)$ is the predicted noise. In the reversed diffusion process, we gradually remove the predicted noise from the current noisy image to recover the image of the last time step $t-1$. Refer to \cite{ho2020denoising} for more Details about the de-noising process.

\subsubsection{Hierarchical Architecture} Inspired by Ronneberger \etal \cite{ronneberger2015u}, who propose to connect the feature maps from encoder and decoder for horizontal feature fusion, we introduce a generic hierarchical architecture to refine and fuse features vertically. Specifically, we assign a building block and a feature decoding block for each stage in the decoder to generate multi-resolution predictions, denoted as $x_1',x_2',\cdots,x_n'$. We introduce a combination layer, where all predictions are weighted with a learnable position encoding and integrated to produce the final output:
\begin{equation}
    x' = \mathcal{P}^{-1}\left( \osum_{i=1}^n(\mathcal{S}(x'_i, x'_n) + p_i)\right).
\end{equation}
In the above, $\osum$ denotes concatenation, $\mathcal{S}(x, y)$ scales $x$ to $y$'s size. Meanwhile, the target $x$ is scaled to the corresponding shapes of predictions to construct a pyramid loss for feature refinements:
\begin{equation}
    \mathcal{L}_\text{pyramid}(x, x'_1, \cdots, x'_n) = \frac{1}{n}\sum_{i=1}^n\mathcal{L}_\text{model}(x'_i,\mathcal{S}(x, x'_i)),
\end{equation}
where $\mathcal{L}_\text{model}$ is the original non-hierarchical loss. The overall loss is computed through the following equation:
\begin{equation}
    \mathcal{L}_\text{H-model} = \mathcal{L}_\text{model}(x, x') + \lambda\mathcal{L}_\text{pyramid}(x, x'_1, \cdots, x'_n)
\end{equation}
The hierarchical architecture is an improvement of our base architecture aiming at discovering and combining local fine details and global structures, both of which are important for high-quality segmentation and reconstruction. Therefore, we derive the hierarchical versions of SeM and AE, denoted as H-SeM and H-AE, respectively. A vertical feature fusion for DDPM is not recommended, because we predict noise instead of reconstructing the image in the reverse diffusion process. Noise usually doesn't contain clear global structures, and scaling the noise map may cause severe loss of details. Fig.~\ref{fig:pyramid} gives an example of a segmentation model constructed with H-SeM and a U-shaped encoder using ConvBlock.

\begin{figure*}[h]
	\centering
	\includegraphics[width=2\columnwidth, angle=0]{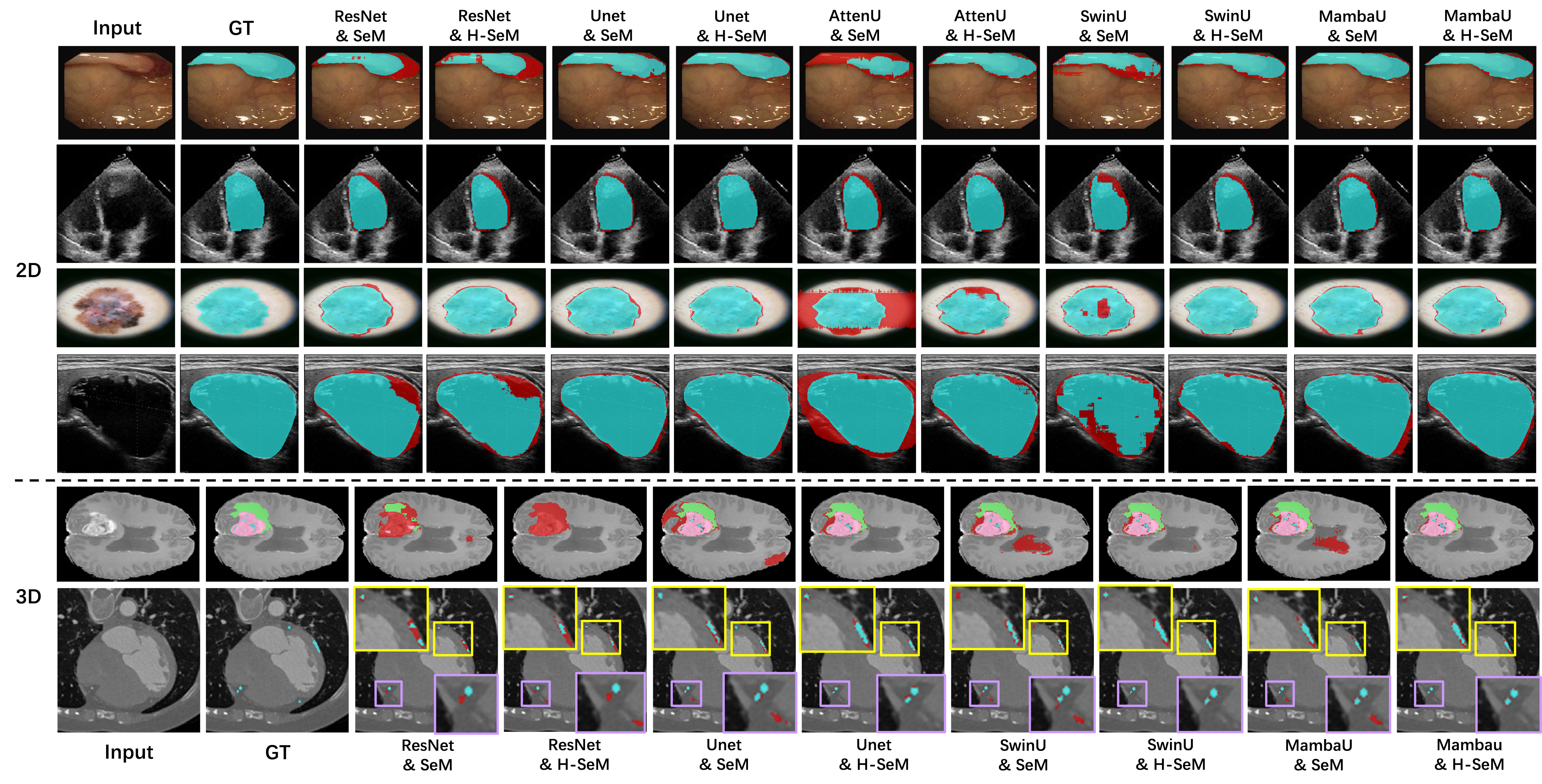}
        \vspace{-0.5em}
	\caption{Quantitative results of segmentation models. The top four rows show segmentation results for 2D image datasets: CVC-ClinicDB, Echonet, ISIC, and TN3K. The bottom two rows show segmentation results of the middle slices for 3D image datasets: BraTS21 and ImageCAS. The pixels highlighted in red represent incorrect predictions.}
	\label{fig:res_seg}
 \vspace{-1em}
\end{figure*}
\section{Experimental Results}
\subsection{Datasets and Evaluation Metrics}
For segmentation, we evaluate our methods on six public datasets. CVC-ClinicDB \cite{bernal2015wm} is a polyp segmentation dataset of 612 images with a resolution from 31 colonoscopy video sequences. Echonet \cite{ouyang2019echonet} includes 10,030 labeled echocardiogram images for chamber segmentation. ISIC \cite{gutman2016skin} was published by the International Skin Imaging Collaboration as a large-scale dataset of 1279 dermoscopy images for lesion segmentation. TN3K \cite{gong2021multi} is an open-access dataset containing 3493 thyroid nodule images with high-quality nodule masks labeling. BraTS21 \cite{baid2021rsna} is a magnetic resonance image (MRI) dataset that provides 2000 3D images with segmentation labels of the different glioma sub-regions. ImageCAS \cite{zeng2023imagecas} contains 1000 3D images for coronary artery segmentation. Images from CVC-ClinicDB, Echonet, ISIC, and TN3K are resized to the dimensions of 320$\times$256, 128$\times$128, 384$\times$256 and 256$\times$256 pixels, respectively. For BraTS21, all images are cropped to the region of non-zero values and resized to a shape of 120$\times$190$\times$120 voxels. For ImageCAS, the images are resized to a shape of 192$\times$192$\times$96 voxels.
We also evaluate reconstruction accuracy on the FastMRI dataset introduced by Zbontar \etal  \cite{zbontar2018fastmri}, which aims to accelerate magnetic resonance imaging by taking fewer measurements. Specifically, we focus on the single-coil knee MRI reconstruction task that contains 1172 cases. Each case contains a k-space raw volume and the corresponding emulated single-coil (ESC) image. We randomly masked 99\% of low-frequency k-space lines, leading to a theoretical 10-fold speedup over fully-sampled single-coil imaging. Zero-filled reconstruction was further performed on the masked k-space volume to obtain the noisy images, which serve as the input of reconstruction models. Similar to \cite{zbontar2018fastmri}, we treat 3D images as multiple 2D slices and all images are cropped to the central 320$\times$320 pixel region to compensate for readout-direction oversampling. The first 5 slices of all samples are discarded due to limited information, resulting in a dataset containing 36,017 images. This dataset is also used for the evaluation of generation models.

For all datasets, we randomly split samples into 5 folds, where the first 3 folds are used for training, and the 4th fold is used for validation. Evaluations are performed on the 5th fold. We compute Dice score and mIoU for segmentation evaluations. PSNR \cite{hore2010image} and SSIM \cite{wang2004image} are used to evaluate reconstruction accuracy. Because there are no universal metrics for evaluating medical image generation, we use the noisy images as input conditions of generation models for high-quality MRI reconstruction so that the results can be evaluated through reconstruction metrics.

\subsection{Experimental Setup}
\subsubsection{Segmentation}We construct segmentation models with SeM architecture and different encoders including ResNet \cite{he2016deep}, U-Net \cite{ronneberger2015u}, CAtten-U \cite{ho2020denoising}, Swin-U \cite{cao2022swin} and Mamba-U \cite{wang2024mamba}, which are constructed with ResConvBlock, ConvBlock, ConvBlock plus MSA, SwinBlock and MambaBlock, respectively. In the above, U-Net, CAtten-U, Swin-U and Mamba-U are U-shaped encoders. The corresponding hierarchical models are constructed using H-SeM. Note that the huge GPU memory requirements prevent us from training models with CAtten-U on 3D image patches. Models trained on 2D and 3D image datasets contain 2 and 3 down-sampling layers, respectively. The number of middle layers is set to 2. We employ a hybrid loss function that combines Dice and cross-entropy loss for all segmentation models, which are trained using Adam \cite{kingma2014adam} optimizer for 500 epochs. The learning rate starts from $3\times10^{-4}$ and follows a linear decaying schedule. 
\subsubsection{Reconstruction}We construct reconstruction models with AE architecture and different encoders including ResNet, U-Net, Atten-U, Swin-U, and Mamba-U. H-AE architecture is used to construct their hierarchical versions. We train all reconstruction models for 50 epochs with L1 loss. Other settings remain the same as segmentation models.
\subsubsection{Generation}We construct generation models with DDPM architecture and use different encoders for noise prediction, which are ResNet, U-Net, Swin-U, and Mamba-U. We use noisy images as input conditions for MRI generation. All models are trained for 200 epochs with L2 loss. The sampling process is accelerated through \cite{song2020denoising}. For a given condition, we integrate 5 or 10 generated samples as the final result, denoted as DDPM (5) and DDPM (10). Other settings remain the same as reconstruction models. 

All experiments about 2D segmentation and reconstruction are conducted on a server with 8 NVIDIA 3090 GPUs. Models for generation and 3D segmentation are trained on 4 NVIDIA A800 GPUs.
\begin{figure*}[ht]
	\centering
	\includegraphics[width=2\columnwidth, angle=0]{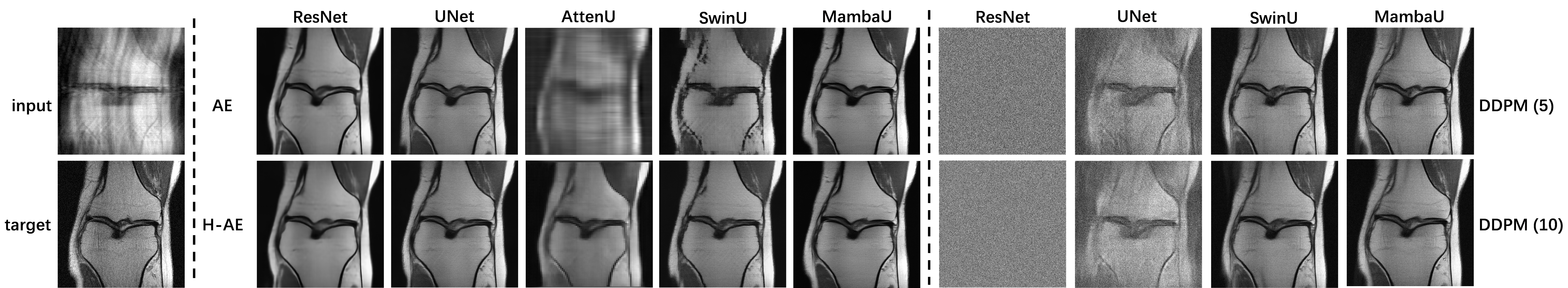}
        \vspace{-0.5em}
	\caption{Quantitative results of reconstruction and generation models for FastMRI dataset.}
	\label{fig:res_ae}
 \vspace{-1em}
\end{figure*}
\begin{table}[h]
\caption{Evaluation of reconstruction (AE, H-AE) and segmentation (DDPM) models.}
\vspace{-2.5em}
\label{tab:recon}
\renewcommand{\arraystretch}{1.1}
\begin{center}
\resizebox{1\columnwidth}{!}{
\begin{tabular}{l|l|cc|l|cc}
\bottomrule
 Encoder   &  Archi  & {PSNR}            & {SSIM}   & Archi  &PSNR &SSIM\\ \hline
\multirow{2}{*}{{ResNet}}   &  AE   & 19.52  & \textit{0.2741} &DDPM (5) & 9.89& 0.0010 \\
                            &  H-AE   & \textit{19.55}  & 0.2715 & DDPM (10) &  \textit{10.14} & \textit{0.0016} \\
\hline
\multirow{2}{*}{{U-Net}}    &  AE   & 20.12  & 0.3429 & DDPM (5) & 11.42& 0.0338 \\
                            &  H-AE   & \textit{20.16}  & \textit{0.3431} & DDPM (10) & \textit{11.69} &\textit{0.0453} \\
\hline
\multirow{2}{*}{{CAtten-U}}    &  AE   & 13.53  & 0.1326  & DDPM (5) & - & - \\
                            &  H-AE   & \textit{16.79}  & \textit{0.1976} & DDPM (10) & - & -\\
\hline
\multirow{2}{*}{{Swin-U}}   &  AE   & 18.44  & 0.2903 &DDPM (5)& 18.32 &0.2721 \\
                            &  H-AE  & \underline{\textit{20.38}}  & \underline{\textit{0.3481}} &DDPM (10)& \underline{\textit{18.65}} & \underline{\textit{0.2940}}  \\
\hline
\multirow{2}{*}{{Mamba-U}}   &  AE  &  20.81 & 0.3648 &DDPM (5)& 19.45 & 0.3093\\
                            &  H-AE  & \textbf{20.99} & \textbf{0.3702} & DDPM (10)& \textbf{19.85}&\textbf{0.3309}\\
\hline

\end{tabular}
}
\end{center}
\vspace{-0.75em}
\end{table}


\subsection{Quantitative and Qualitative Results}
\label{sec: qresults}
Table.~\ref{tab:seg} shows the quantitative results for segmentation models. From the encoder's point of view, Mamba-based methods show the highest superiority. Mamba-U consistently ranks in the top 2 across all six segmentation datasets and achieves the highest accuracy on three of them. U-Net also demonstrates strong performance, securing a top-2 ranking on four datasets and achieving the highest accuracy on two of them. Swin-U achieves top-2 performance on two datasets. ResNet tends to overfit, highlighting that the U-shaped design significantly enhances the robustness of the segmentation models. CAtten-U has the poorest performance, suggesting that a straightforward combination of convolution and attention mechanisms may not be optimal for medical segmentation. Moreover, this approach severely increases the computational and memory burden. From the model architecture's perspective, our hierarchical design improves performance across all U-shaped networks, which demonstrates the effectiveness of the proposed vertical feature fusion. A comprehensive comparison of qualitative results is provided in Fig.~\ref{fig:res_seg}.

Table.~\ref{tab:recon} shows the quantitative results for reconstruction models and generation models. All U-shaped reconstruction models benefit from our hierarchical architecture. When using conditional DDPM for medical image restoration, a larger ensemble number leads to higher accuracy. Similar to segmentation models, Mamba-U achieves the highest reconstruction accuracy. Convolution-based encoders also have satisfactory capability for reconstruction, although they perform poorly in terms of generation. We notice that sequence modeling-based encoders generally outperform convolution-based encoders for reconstruction and generation, indicating that exploring long-range dependency is particularly effective for high-frequency component prediction. Qualitative results are illustrated in Fig.~\ref{fig:res_ae}.
\begin{table}[h]
\caption{Comparison of Training time and GPU memory consumption for networks with different building blocks. }
\vspace{-2.5em}
\label{tab:time_space}
\begin{center}
\resizebox{\columnwidth}{!}{
\renewcommand{\arraystretch}{1.1}
\begin{tabular}{c|c|c|cccc}
\bottomrule
  Building  & \multirow{2}{*}{{\#Layer}}    &  \multirow{2}{*}{{Patch}}  & \#Params  &   Memory    & Training    & Inference\\
    block       &                                 &                           
               &  (M)     & (GB)         & time (s) &time (s)  \\
\hline
   \multirow{8}{*}{{ConvBlock}} &  \multirow{4}{*}{{16}}    & 64$^2$   &                               & 0.056  & 8.69 & 0.73\\
                                &                         & 128$^2$   & \multirow{-2}{*}{{3.88}}  & 0.22  & 8.75 & 0.74 \\
                                &                         & 64$^3$   &                        & 0.62  & 9.16 & 1.45\\
                                &                         & 128$^3$   & \multirow{-2}{*}{{12.62}} & 2.95  & 30.77 & 9.31\\
\cdashline{2-7}
                                &  \multirow{4}{*}{{32}}    & 64$^2$   &                         &  0.16  & 14.17 & 1.23\\
                                &                         & 128$^2$   & \multirow{-2}{*}{{6.98}}  &  0.20 & 14.52 & 1.24\\
                                &                         & 64$^3$   &                            & 0.77  &14.94 & 2.47\\
                                &                         & 128$^3$   & \multirow{-2}{*}{{21.92}} & 4.24  & 53.26 & 15.81\\
\hline
  \multirow{8}{*}{{SwinBlock}} &  \multirow{4}{*}{{16}}    & 64$^2$   &                             & 0.29  & 26.76 & 4.02\\
                                &                         & 128$^2$   &  \multirow{-2}{*}{{4.38}} & 0.54  & 27.95 & 4.14\\
                                &                         & 64$^3$   &               & 4.97  &45.41 & 15.54\\
                                &                         & 128$^3$   &  \multirow{-2}{*}{{4.75}} & 38.53  & 332.42 & 120.16\\
\cdashline{2-7}
                                &  \multirow{4}{*}{{32}}    & 64$^2$   &                         & 0.41 & 33.52 & 7.49\\
                                &                         & 128$^2$   & \multirow{-2}{*}{{8.53}} &  0.86 & 33.91 & 7.73\\
                                &                         & 64$^3$   &                       & 8.32  &87.49 & 30.45\\
                                &                         & 128$^3$   & \multirow{-2}{*}{{8.53}} & 65.41  & 629.67 & 230.64\\
\hline
  \multirow{8}{*}{{MambaBlock}} &  \multirow{4}{*}{{16}}    & 64$^2$   &                         & 0.34  & 28.13 & 4.55\\
                                &                         & 128$^2$   & \multirow{-2}{*}{{5.53}} & 0.73  &27.62 & 4.53\\
                                &                         & 64$^3$   &                       & 4.35  & 60.28 & 17.55\\
                                &                         & 128$^3$   & \multirow{-2}{*}{{5.86}} &  34.21  & 505.96 & 149.46\\
\cdashline{2-7}
                                &  \multirow{4}{*}{{32}}    & 64$^2$   &                         & 0.50  & 35.02 & 8.77\\
                                &                         & 128$^2$   & \multirow{-2}{*}{{10.85}} &  1.24 & 35.38 & 8.80\\
                                &                         & 64$^3$   &                       &  7.11  &110.08 & 32.44\\
                                &                         & 128$^3$   & \multirow{-2}{*}{{11.24}} & 55.73  & 920.41 & 271.75\\
\hline
\end{tabular}
}
\end{center}
\vspace{-1.25em}
\end{table}
\subsection{Time and Memory Consumption}
We also compare the training time and GPU memory consumption of different building blocks as shown in Table.~\ref{tab:time_space}. We construct models with SeM architecture and encoders using building blocks ConvBlock, SwinBlock, and MambaBlock. Each model contains two down-sampling and two up-sampling layers. The batch size is set to 2. We measure the runtime and reserved GPU memory for 500 training and inference iterations on a single A800 GPU. 
As the network depth increases, the number of parameters, GPU memory consumption, and training/inference time for all models increase linearly. While ConvBlock achieves the highest time and memory efficiency, gaps between other building blocks and ConvBlock are acceptable for 2D images. When we switch the input to 3D images, SwinBlock and MambaBlock do not experience a significant increase in the number of parameters because they process images as sequences. However, the length of sequence for 3D images increases cubically, which leads to a sudden leap in runtime and memory consumption. As the image size increases, the advantages of ConvBlock on time and space complexity become more significant. The training/inference time of models using ConvBlock is 5.26$\times$ and 6.97$\times$ faster than models using SwinBlock and MambaBlock for input patches with a size of 64$^3$. When the patch size grows to 128$^3$, it is 11.3$\times$ and 16.86$\times$ faster to train a CNN compared to training models with SwinBlock and MambaBlock under the same experimental settings. In practice, the differences in training duration are more pronounced because CNNs can be trained with much larger batch sizes. Due to the prevalence of 3D data in medical images and the superior performance of CNNs as analyzed in the above and Sec.~\ref{sec: qresults}, we believe that CNNs will continue to hold an irreplaceable position in medical image segmentation and reconstruction.

\section{Conclusion, Discussion, and Future Works}

In this paper, we present Flemme, a general learning platform aiming at the rapid and flexible development of deep learning models for medical images. We introduce various encoders based on convolution, transformer, and SSM backbones, combining SeM, AE, and DDPM architectures to facilitate model creation for medical image segmentation, reconstruction, and generation. We also implement H-SeM and H-AE by employing a generic hierarchical architecture for vertical feature refinement and fusion. Extensive experiments on multiple datasets with multiple modalities showcase that this design improves model performance across different encoders. 

Benefiting from the advantages of our platform in model creation and hyper-parameter control, we conduct a fair comparison and analysis of various encoders regarding runtime, memory consumption, and performance across different tasks. We confirm that CNNs are still playing essential roles in medical image segmentation and reconstruction, although we notice that there are notable limitations of convolution compared to sequence-modeling backbones in generation. Given these findings, we recommend SSM-based networks as the optimal choice when sufficient computational resources are available, especially for high-frequency component prediction.

For future works, we are actively working on expanding our platform to handle diverse data types such as modeling point cloud and graph, whose importance in biomedical applications is becoming increasingly evident. Additionally, high-quality medical image generation, as well as automated quality assessment of generated medical images, will also be the focus of our future research. 

\section*{Acknowledgment}
This work is supported by the Natural Science Foundation of China (Grant 62371270), the Major Key Project of PCL (Grant PCL2023A09, Pengcheng Laboratory), and Shenzhen Key Laboratory of Ubiquitous Data Enabling (No.ZDSYS20220527171406015).

\bibliographystyle{IEEEtran}
\bibliography{bib}

\begin{thebibliography}{10}
\providecommand{\url}[1]{#1}
\csname url@samestyle\endcsname
\providecommand{\newblock}{\relax}
\providecommand{\bibinfo}[2]{#2}
\providecommand{\BIBentrySTDinterwordspacing}{\spaceskip=0pt\relax}
\providecommand{\BIBentryALTinterwordstretchfactor}{4}
\providecommand{\BIBentryALTinterwordspacing}{\spaceskip=\fontdimen2\font plus
\BIBentryALTinterwordstretchfactor\fontdimen3\font minus \fontdimen4\font\relax}
\providecommand{\BIBforeignlanguage}[2]{{%
\expandafter\ifx\csname l@#1\endcsname\relax
\typeout{** WARNING: IEEEtran.bst: No hyphenation pattern has been}%
\typeout{** loaded for the language `#1'. Using the pattern for}%
\typeout{** the default language instead.}%
\else
\language=\csname l@#1\endcsname
\fi
#2}}
\providecommand{\BIBdecl}{\relax}
\BIBdecl

\bibitem{krizhevsky2012imagenet}
A.~Krizhevsky, I.~Sutskever, and G.~E. Hinton, ``Imagenet classification with deep convolutional neural networks,'' \emph{Advances in neural information processing systems}, vol.~25, 2012.

\bibitem{deng2009imagenet}
J.~Deng, W.~Dong, R.~Socher, L.-J. Li, K.~Li, and L.~Fei-Fei, ``Imagenet: A large-scale hierarchical image database,'' in \emph{2009 IEEE conference on computer vision and pattern recognition}.\hskip 1em plus 0.5em minus 0.4em\relax Ieee, 2009, pp. 248--255.

\bibitem{ronneberger2015u}
O.~Ronneberger, P.~Fischer, and T.~Brox, ``U-net: Convolutional networks for biomedical image segmentation,'' in \emph{Medical image computing and computer-assisted intervention--MICCAI 2015: 18th international conference, Munich, Germany, October 5-9, 2015, proceedings, part III 18}.\hskip 1em plus 0.5em minus 0.4em\relax Springer, 2015, pp. 234--241.

\bibitem{vaswani2017attention}
A.~Vaswani, N.~Shazeer, N.~Parmar, J.~Uszkoreit, L.~Jones, A.~N. Gomez, {\L}.~Kaiser, and I.~Polosukhin, ``Attention is all you need,'' \emph{Advances in neural information processing systems}, vol.~30, 2017.

\bibitem{devlin2018bert}
J.~Devlin, M.-W. Chang, K.~Lee, and K.~Toutanova, ``Bert: Pre-training of deep bidirectional transformers for language understanding,'' \emph{arXiv preprint arXiv:1810.04805}, 2018.

\bibitem{dosovitskiy2020image}
A.~Dosovitskiy, L.~Beyer, A.~Kolesnikov, D.~Weissenborn, X.~Zhai, T.~Unterthiner, M.~Dehghani, M.~Minderer, G.~Heigold, S.~Gelly \emph{et~al.}, ``An image is worth 16x16 words: Transformers for image recognition at scale,'' \emph{arXiv preprint arXiv:2010.11929}, 2020.

\bibitem{he2016deep}
K.~He, X.~Zhang, S.~Ren, and J.~Sun, ``Deep residual learning for image recognition,'' in \emph{Proceedings of the IEEE conference on computer vision and pattern recognition}, 2016, pp. 770--778.

\bibitem{liu2021swin}
Z.~Liu, Y.~Lin, Y.~Cao, H.~Hu, Y.~Wei, Z.~Zhang, S.~Lin, and B.~Guo, ``Swin transformer: Hierarchical vision transformer using shifted windows,'' in \emph{Proceedings of the IEEE/CVF international conference on computer vision}, 2021, pp. 10\,012--10\,022.

\bibitem{gu2023mamba}
A.~Gu and T.~Dao, ``Mamba: Linear-time sequence modeling with selective state spaces,'' \emph{arXiv preprint arXiv:2312.00752}, 2023.

\bibitem{zhu2024vision}
L.~Zhu, B.~Liao, Q.~Zhang, X.~Wang, W.~Liu, and X.~Wang, ``Vision mamba: Efficient visual representation learning with bidirectional state space model,'' \emph{arXiv preprint arXiv:2401.09417}, 2024.

\bibitem{liu2024vmambavisualstatespace}
Y.~Liu, Y.~Tian, Y.~Zhao, H.~Yu, L.~Xie, Y.~Wang, Q.~Ye, and Y.~Liu, ``Vmamba: Visual state space model,'' 2024.

\bibitem{liu2022convnet}
Z.~Liu, H.~Mao, C.-Y. Wu, C.~Feichtenhofer, T.~Darrell, and S.~Xie, ``A convnet for the 2020s,'' in \emph{Proceedings of the IEEE/CVF conference on computer vision and pattern recognition}, 2022, pp. 11\,976--11\,986.

\bibitem{kingma2013auto}
D.~P. Kingma and M.~Welling, ``Auto-encoding variational bayes,'' \emph{arXiv preprint arXiv:1312.6114}, 2013.

\bibitem{ho2020denoising}
J.~Ho, A.~Jain, and P.~Abbeel, ``Denoising diffusion probabilistic models,'' \emph{Advances in neural information processing systems}, vol.~33, pp. 6840--6851, 2020.

\bibitem{song2020denoising}
J.~Song, C.~Meng, and S.~Ermon, ``Denoising diffusion implicit models,'' \emph{arXiv preprint arXiv:2010.02502}, 2020.

\bibitem{wu2018group}
Y.~Wu and K.~He, ``Group normalization,'' in \emph{Proceedings of the European conference on computer vision (ECCV)}, 2018, pp. 3--19.

\bibitem{grathwohl2018ffjord}
W.~Grathwohl, R.~T. Chen, J.~Bettencourt, I.~Sutskever, and D.~Duvenaud, ``Ffjord: Free-form continuous dynamics for scalable reversible generative models,'' \emph{arXiv preprint arXiv:1810.01367}, 2018.

\bibitem{ba2016layer}
J.~L. Ba, J.~R. Kiros, and G.~E. Hinton, ``Layer normalization,'' \emph{arXiv preprint arXiv:1607.06450}, 2016.

\bibitem{larsson2016fractalnet}
G.~Larsson, M.~Maire, and G.~Shakhnarovich, ``Fractalnet: Ultra-deep neural networks without residuals,'' \emph{arXiv preprint arXiv:1605.07648}, 2016.

\bibitem{dao2024transformers}
T.~Dao and A.~Gu, ``Transformers are ssms: Generalized models and efficient algorithms through structured state space duality,'' \emph{arXiv preprint arXiv:2405.21060}, 2024.

\bibitem{cao2022swin}
H.~Cao, Y.~Wang, J.~Chen, D.~Jiang, X.~Zhang, Q.~Tian, and M.~Wang, ``Swin-unet: Unet-like pure transformer for medical image segmentation,'' in \emph{European conference on computer vision}.\hskip 1em plus 0.5em minus 0.4em\relax Springer, 2022, pp. 205--218.

\bibitem{chen2021transunet}
J.~Chen, Y.~Lu, Q.~Yu, X.~Luo, E.~Adeli, Y.~Wang, L.~Lu, A.~L. Yuille, and Y.~Zhou, ``Transunet: Transformers make strong encoders for medical image segmentation,'' \emph{arXiv preprint arXiv:2102.04306}, 2021.

\bibitem{wang2024mamba}
Z.~Wang, J.-Q. Zheng, Y.~Zhang, G.~Cui, and L.~Li, ``Mamba-unet: Unet-like pure visual mamba for medical image segmentation,'' \emph{arXiv preprint arXiv:2402.05079}, 2024.

\bibitem{van2016conditional}
A.~Van~den Oord, N.~Kalchbrenner, L.~Espeholt, O.~Vinyals, A.~Graves \emph{et~al.}, ``Conditional image generation with pixelcnn decoders,'' \emph{Advances in neural information processing systems}, vol.~29, 2016.

\bibitem{ho2022classifier}
J.~Ho and T.~Salimans, ``Classifier-free diffusion guidance,'' \emph{arXiv preprint arXiv:2207.12598}, 2022.

\bibitem{bernal2015wm}
J.~Bernal, F.~J. S{\'a}nchez, G.~Fern{\'a}ndez-Esparrach, D.~Gil, C.~Rodr{\'\i}guez, and F.~Vilari{\~n}o, ``Wm-dova maps for accurate polyp highlighting in colonoscopy: Validation vs. saliency maps from physicians,'' \emph{Computerized medical imaging and graphics}, vol.~43, pp. 99--111, 2015.

\bibitem{ouyang2019echonet}
D.~Ouyang, B.~He, A.~Ghorbani, M.~P. Lungren, E.~A. Ashley, D.~H. Liang, and J.~Y. Zou, ``Echonet-dynamic: a large new cardiac motion video data resource for medical machine learning,'' in \emph{NeurIPS ML4H Workshop}, 2019, pp. 1--11.

\bibitem{gutman2016skin}
D.~Gutman, N.~C. Codella, E.~Celebi, B.~Helba, M.~Marchetti, N.~Mishra, and A.~Halpern, ``Skin lesion analysis toward melanoma detection: A challenge at the international symposium on biomedical imaging (isbi) 2016, hosted by the international skin imaging collaboration (isic),'' \emph{arXiv preprint arXiv:1605.01397}, 2016.

\bibitem{gong2021multi}
H.~Gong, G.~Chen, R.~Wang, X.~Xie, M.~Mao, Y.~Yu, F.~Chen, and G.~Li, ``Multi-task learning for thyroid nodule segmentation with thyroid region prior,'' in \emph{2021 IEEE 18th international symposium on biomedical imaging (ISBI)}.\hskip 1em plus 0.5em minus 0.4em\relax IEEE, 2021, pp. 257--261.

\bibitem{baid2021rsna}
U.~Baid, S.~Ghodasara, S.~Mohan, M.~Bilello, E.~Calabrese, E.~Colak, K.~Farahani, J.~Kalpathy-Cramer, F.~C. Kitamura, S.~Pati \emph{et~al.}, ``The rsna-asnr-miccai brats 2021 benchmark on brain tumor segmentation and radiogenomic classification,'' \emph{arXiv preprint arXiv:2107.02314}, 2021.

\bibitem{zeng2023imagecas}
A.~Zeng, C.~Wu, G.~Lin, W.~Xie, J.~Hong, M.~Huang, J.~Zhuang, S.~Bi, D.~Pan, N.~Ullah \emph{et~al.}, ``Imagecas: A large-scale dataset and benchmark for coronary artery segmentation based on computed tomography angiography images,'' \emph{Computerized Medical Imaging and Graphics}, vol. 109, p. 102287, 2023.

\bibitem{zbontar2018fastmri}
J.~Zbontar, F.~Knoll, A.~Sriram, T.~Murrell, Z.~Huang, M.~J. Muckley, A.~Defazio, R.~Stern, P.~Johnson, M.~Bruno \emph{et~al.}, ``fastmri: An open dataset and benchmarks for accelerated mri,'' \emph{arXiv preprint arXiv:1811.08839}, 2018.

\bibitem{hore2010image}
A.~Hore and D.~Ziou, ``Image quality metrics: Psnr vs. ssim,'' in \emph{2010 20th international conference on pattern recognition}.\hskip 1em plus 0.5em minus 0.4em\relax IEEE, 2010, pp. 2366--2369.

\bibitem{wang2004image}
Z.~Wang, A.~C. Bovik, H.~R. Sheikh, and E.~P. Simoncelli, ``Image quality assessment: from error visibility to structural similarity,'' \emph{IEEE transactions on image processing}, vol.~13, no.~4, pp. 600--612, 2004.

\bibitem{kingma2014adam}
D.~P. Kingma and J.~Ba, ``Adam: A method for stochastic optimization,'' \emph{arXiv preprint arXiv:1412.6980}, 2014.

\end{thebibliography}

\end{document}